\documentclass[epj,twocolumn]{webofc}

\usepackage[varg]{txfonts}   

\usepackage{color}
\usepackage{graphicx}
\usepackage{amssymb,amsmath,enumerate}
\usepackage[colorlinks=true,linkcolor=blue,citecolor=blue,urlcolor=blue]{hyperref}

\woctitle{Subnuclear Structure of Matter: Achievements and Challenges}
\begin{document}
\title{Baryons with functional methods}
%
%


\author{\firstname{Christian~S.} \lastname{Fischer}\inst{1,2}\fnsep\thanks{\email{christian.fischer@physik.uni-giessen.de}}
}

\institute{Institut f\"ur Theoretische Physik, Justus-Liebig--Universit\"at Giessen, 35392 Giessen, Germany
\and
           HIC for FAIR Giessen, 35392 Giessen, Germany
          }

\abstract{%
We summarise recent results on the spectrum of ground-state and excited baryons and
their form factors in the framework of functional methods.
As an improvement upon similar approaches
we explicitly take into account the underlying momentum-dependent dynamics
of the quark-gluon interaction that leads to dynamical chiral symmetry
breaking. For light octet and decuplet baryons we find a spectrum in very
good agreement with experiment, including the level ordering between the
positive- and negative-parity nucleon states. Comparing the three-body framework
with the quark-diquark approximation, we do not find significant differences in
the spectrum for those states that have been calculated in both frameworks.
This situation is different in the electromagnetic form factor of the $\Delta$,
which may serve to distinguish both pictures by comparison with experiment
and lattice QCD.
}

\maketitle
\section{Introduction}\label{intro}

In these proceedings we summarise our progress on the properties of baryons
as obtained from the functional approach (Dyson-Schwinger and Bethe-Salpeter equations)
to QCD. All topics discussed in this contribution are covered in much more detail in a very recent
review on baryons as covariant three-quark bound states~\cite{Eichmann:2016yit}.

Understanding the baryon excitation spectrum of QCD and the internal structure of baryons
using electromagnetic and other probes is one of the key elements in unravelling the
structure of the strong interaction. In the past years, significant experimental progress
has been made by the analysis of data from photo- and electroproduction experiments at
JLAB, ELSA and MAMI~\cite{Aznauryan:2009mx,Aznauryan:2012ba,Sarantsev:2007aa}. As a result,
a number of new radial and orbital excitations of the ground-state baryons have been
added to the PDG~\cite{Agashe:2014kda}.

On the theoretical side, a number of non-perturbative methods contribute a number of interesting aspects
to our understanding of baryons, which are
not always in agreement with each other. In general, the quark model still serves as a possible
standard by which one may define and distinguish the `expected' from the `unexpected'. Its
discussion along the years has created a list of standard problems in baryon physics~\cite{Capstick:2000qj,Klempt:2009pi}.
Amongst these are the problem of missing resonances and,
related, the question whether baryons should better be described as quark-diquark rather than
three-quark states, thus reducing the number of possible internal excitations. On the other hand,
dynamical coupled-channel calculations have suggested that parts of the spectrum may even be
generated purely dynamically, i.e.\ without a `bare' three-quark or quark-diquark
seed~\cite{Doring:2009yv,Bruns:2010sv}. Lattice QCD contributes its own share with fairly small
error bars and very good agreement between different lattice groups and approaches for most ground-state
masses. Differences, however, occur for the excited states, where a number of technical
subtleties still obscure the picture.

In the past years we explored another non-perturbative approach to QCD, namely the functional
methods of Dyson-Schwinger and Bethe-Salpeter equations (DSEs and BSEs). They can be used on different
levels of sophistication, ranging from NJL-like truncations with contact interactions, quark-diquark
models using ans\"atze for propagators and wave functions as input, rainbow-ladder truncations
using effective quark-gluon interactions, to beyond-rainbow-ladder approaches explicitly solving
a range of underlying DSEs for the QCD Green's functions; see~\cite{Eichmann:2016yit} for an overview.
In the work summarised here we have used the rainbow-ladder truncation as a minimal baseline
but also explored effects beyond rainbow-ladder. In addition, we do not rely on the widely used
quark-diquark approximation but provided first solutions for the masses of ground and selected
excited states as well as baryon form factors from the three-body Faddeev equation. We are therefore
in a position to be able to systematically compare both approaches using the same underlying assumptions for
the quark-gluon interaction.

           \begin{figure*}[t]
            \centering
            \includegraphics[width=0.85\textwidth]{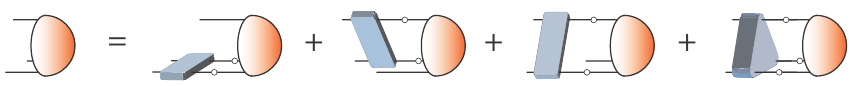}
            \caption{Three-quark Faddeev equation.}
            \label{fig:faddeev}
            \includegraphics[width=0.9\textwidth]{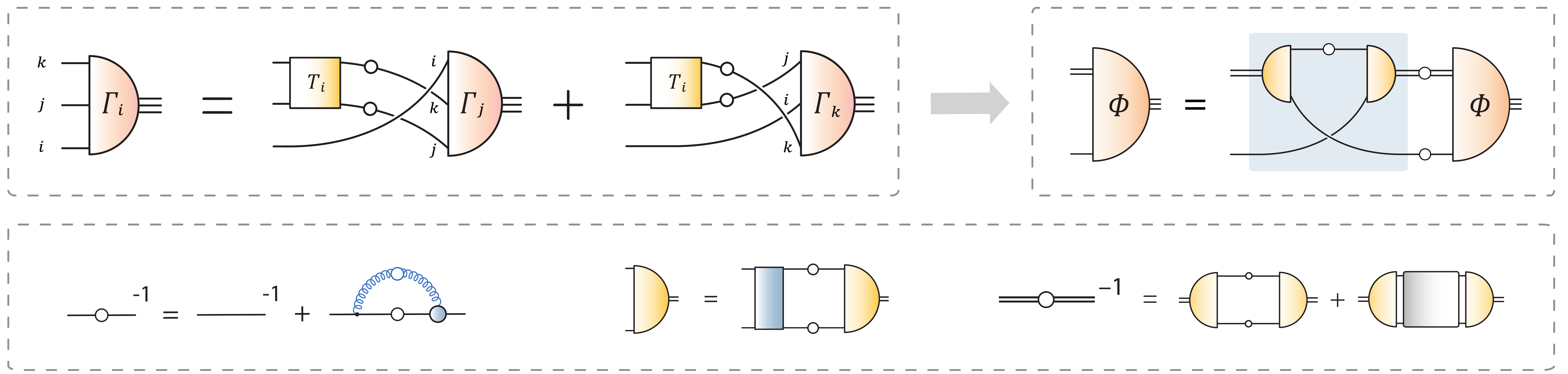}
            \caption{Simplification of the Faddeev equation in Fig.~\ref{fig:faddeev}
            (\textit{top left}) to the quark-diquark Bethe-Salpeter equation (\textit{top right}).
                     The bottom panel shows the ingredients that enter in the equation and are calculated
                     beforehand: the quark propagator, diquark Bethe-Salpeter amplitudes and diquark propagators.}
            \label{fig:quark-diquark}
            \end{figure*}

Our results for the light baryon spectrum will be discussed in Sec.~\ref{sec-3}. Corresponding
results for form factors will be the topic of Sec.~\ref{sec-4}, with a focus on the $\Delta$ electromagnetic form factors.
First, however, we outline the framework in the next section.

\section{The functional approach}\label{sec-2}

In functional frameworks the masses and wave functions of baryons are extracted from their gauge-invariant
poles in the (gauge-dependent) quark six-point Green function. This procedure is related to the corresponding
one in lattice QCD, see Ref.~\cite{Eichmann:2016yit} for more detailed explanations. An exact equation that can
be derived in this approach is the covariant three-body Faddeev equation in Fig.~\ref{fig:faddeev}.
Its ingredients are the fully dressed quark propagator (solid line with open circle) as well as the quark two-body
and irreducible three-body interactions. The latter ones have been neglected so far. The equation determines the baryon's
Faddeev amplitude (the shaded half-sphere) and the baryon bound-state mass.

A considerable simplification is obtained once the quark-diquark approximation is chosen. To this end,
one expands the quark-quark scattering matrix in terms of separable diquark contributions, resulting
in the quark-diquark Bethe-Salpeter equation in the top-right corner of Fig.~\ref{fig:quark-diquark}.
Its main ingredients are again the dressed quark propagator, the diquark propagator (double line with
open circle) as well as the diquark Bethe-Salpeter amplitude. In turn, the latter needs to be
calculated from the diquark Bethe-Salpeter equation (bottom center diagram of Fig.~\ref{fig:quark-diquark}).
The dressed quark propagator is determined from its Dyson-Schwinger equation (bottom left diagram
of Fig.~\ref{fig:quark-diquark}).

An important question in connection with the diquark approximation is how many diquarks to consider.
It turns out that the diquarks with smallest mass are the scalar and axialvector ones~\cite{Maris:2002yu}. While both
of these contribute to the nucleon, the $\Delta$ baryon can only be made of an axialvector diquark due
to the symmetry properties of the diquark wave functions. Thus, scalar and axialvector diquarks together
constitute a minimal set of diquarks necessary to consider. However, as it turns out, the parity partners
of the nucleon and the the $\Delta$ cannot be described adequately without including the heavier
pseudoscalar and vector diquarks in addition~\cite{Eichmann:2016jqx}. Thus in the diquark calculation of the spectrum discussed
below all four types of diquarks have been taken into account.

A systematic comparison between the three-body and diquark-quark approach is only possible once the same
underlying quark-gluon interaction is chosen~\cite{Eichmann:2016hgl}. This distinguishes a quark-diquark approach
based on the dynamics of the underlying QCD from quark-diquark models~\cite{Oettel:1998bk,Oettel:2000jj,Segovia:2015hra}.
In our case we choose the effective interaction introduced by Maris and Tandy in Ref.~\cite{Maris:1999nt};
details are discussed e.g. in~\cite{Eichmann:2016hgl,review}. In this rainbow-ladder framework the two-body interaction
kernel between two quarks is approximated by the exchange of an effective gluon, which contains the gluon
propagator supplemented by those effects of the quark-gluon vertex that can be captured by a function depending
on the gluon momentum only.

\begin{figure*}[t]
        \centering
        \includegraphics[width=0.95\columnwidth]{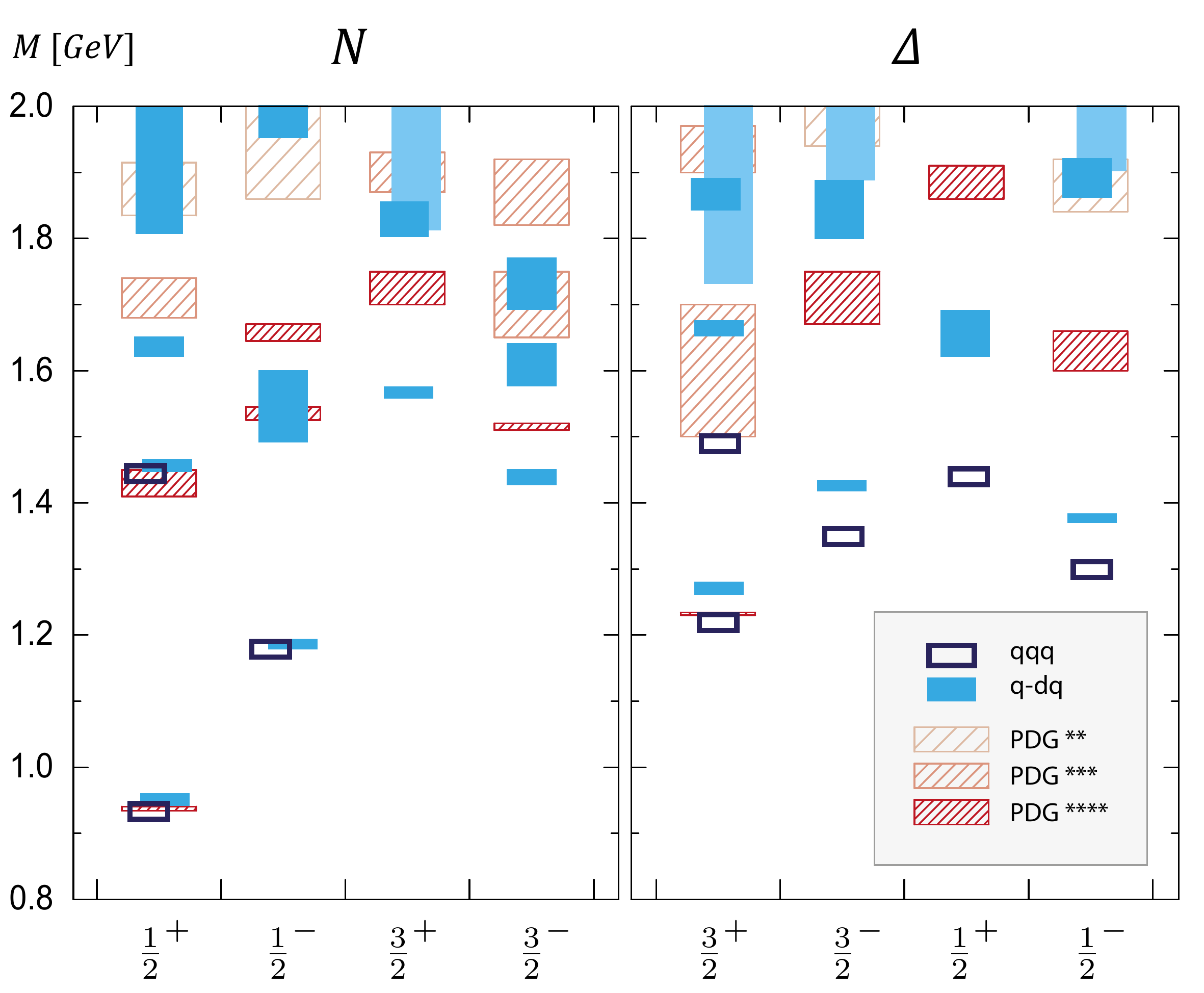}\hfill
        \includegraphics[width=0.95\columnwidth]{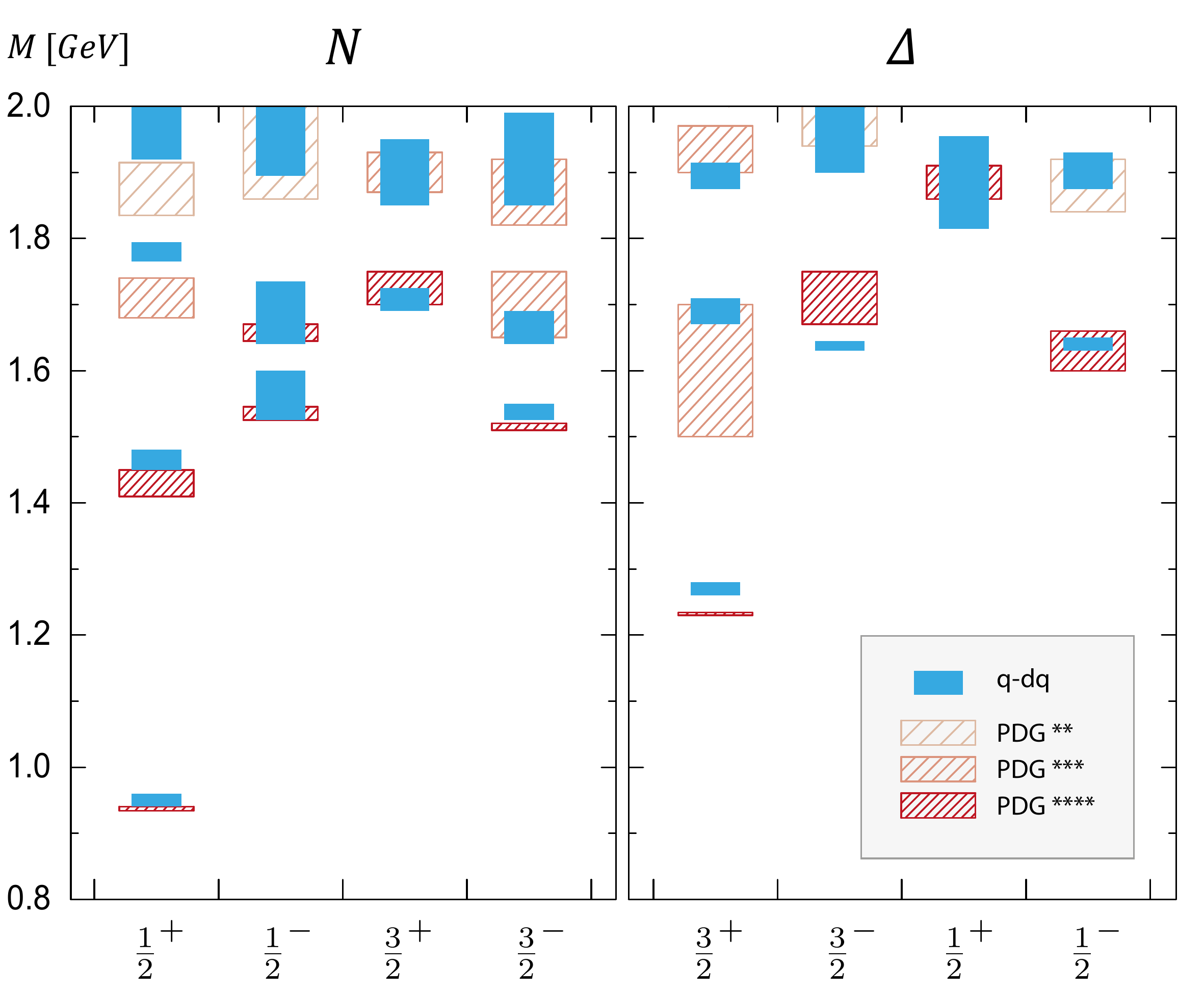}
        \caption{\textit{Left:} Nucleon and $\Delta$ baryon spectrum for $J^P=1/2^\pm$ and $3/2^\pm$ states determined within rainbow-ladder.
                 The three-body results (open boxes) are compared to the quark-diquark spectrum with full diquark content (filled boxes),
                 together with the PDG values and
                 their experimental uncertainties~\cite{Agashe:2014kda}.
                 The widths represent an estimate of the systematic error of our results based on the extrapolated eigenvalue curves
                 of the BSE, see \cite{Eichmann:2016hgl} for details.
                \textit{Right:} Nucleon and $\Delta$ spectrum with reduced strength in the
                                pseudoscalar and vector diquark channels; see text for a detailed discussion.} \label{spectrum}
\end{figure*}

In the rainbow-ladder approximation one arrives at an interaction entirely of vector-vector type. In the heavy quark sector
it is known that such types of interactions are not sufficient to account for the different effects due to spin-spin
and spin-orbit dependent forces between the quarks, and hence cannot describe all meson channels sufficiently well.
In practice, it turns out that the meson spectrum in the pseudoscalar, vector and some tensor channels are well
represented in rainbow-ladder, while others suffer from too much binding~\cite{Hilger:2014nma,Fischer:2014cfa}.
In~\cite{Roberts:2011cf,Eichmann:2016hgl} a simple procedure has been devised to remedy this problem within the quark-diquark approach.
Scalar and axialvector diquarks are well represented in rainbow-ladder, whereas the pseudoscalar and vector
ones are too light. Reducing the interaction strength in these channels by a common factor gauged by the
$\rho-a_1$ splitting remedies this problem.

\section{Light baryon spectrum}\label{sec-3}

In Fig.~\ref{spectrum} we show the resulting spectra from the three-body calculation (left diagram, open boxes), the quark-diquark approximation without correction
(left diagram, filled boxes) and the quark-diquark approximation with improved pseudoscalar and vector diquarks (right diagram, filled boxes)~\cite{Eichmann:2016hgl}
and compare them with the two-, three- and four-star states given by the PDG~\cite{Agashe:2014kda}.

Let us first discuss the nucleon channel. In the left diagram we observe good agreement of the three-body with the quark-diquark
approach for the ground-state nucleon as well as the first radially excited state. The mass of the latter agrees well with
the Breit-Wigner mass of the Roper shown as a red shaded box\footnote{The mass evolution of the Roper with varying pion mass
is discussed in~\cite{Eichmann:2016hgl} and compared with results from lattice QCD.}.
The next two excited states are (numerically) only accessible in the quark-diquark
framework and lie in the same ballpark as the PDG's N(1710) and N(1880). In the nucleon channel with negative parity, both approaches
still agree very well but the states are generally much too low. Clearly, this reflects the above discussed deficiency of the
rainbow-ladder approximation.

A similar picture is observed in the various $\Delta$ channels. The quantitative agreement
between the quark-diquark and the three-body approach is somewhat less pronounced than in the nucleon case but still satisfied
on a semi-quantitative level. Taken at face value, this could mean that interactions beyond the diquark approximation could be more important
for $\Delta$ states than for baryons of nucleon type. This hypothesis needs to be tested in the future.
Again, the (s-wave) $\Delta$-channel delivers results for the ground and first radially excited state in agreement with experiment,
while its parity partners and the other two channels are poorly represented.

The picture changes when the deficiencies of the rainbow-ladder approximation are accounted for by adjusting the binding in the
pseudoscalar and vector diquark channels. As explained above, this is done by introducing one parameter that is adjusted
by the $\rho-a_1$ splitting; everything else is unchanged. The influence of this adjustment depends on the quantum numbers of the baryon
due to different weights of the corresponding diquark contributions as compared to the scalar and axialvector one. It is therefore highly
non-trivial that the states arrange themselves in the one-to-one correspondence with the experimental ones as observed in the right
diagram of Fig.~\ref{spectrum}. Without further corrections, the level ordering between the positive and negative nucleon parity
sector comes out correctly on a quantitative level. Not only the first radial excitations but also the second and third ones are close
to experimental states. We therefore arrive at a consistent and quantitative description of the light baryon spectrum below 2 GeV.

Of course this does not mean that beyond rainbow-ladder effects, like pion-cloud and coupled-channel effects, or
further non-Abelian corrections on the gluon level are absent~\cite{Williams:2015cvx}. Our results merely suggest that these effects are opposite
in sign and largely cancel each other in the nucleon and $\Delta$ channels, whereas in the other channels there is a net effect that can be
absorbed by the extra parameter for the pseudoscalar and vector diquarks that we introduced. In fact, in the meson sector there
are strong indications that cancellations between non-Abelian corrections~\cite{Fischer:2009jm} and pion cloud effects~\cite{Fischer:2008wy}
occur. Both effects have already been explored for baryons on an exploratory level~\cite{Sanchis-Alepuz:2014wea,Sanchis-Alepuz:2015qra}
and work will continue in this direction.

        \begin{figure*}[t]
        \centering
\includegraphics[width=0.95\textwidth]{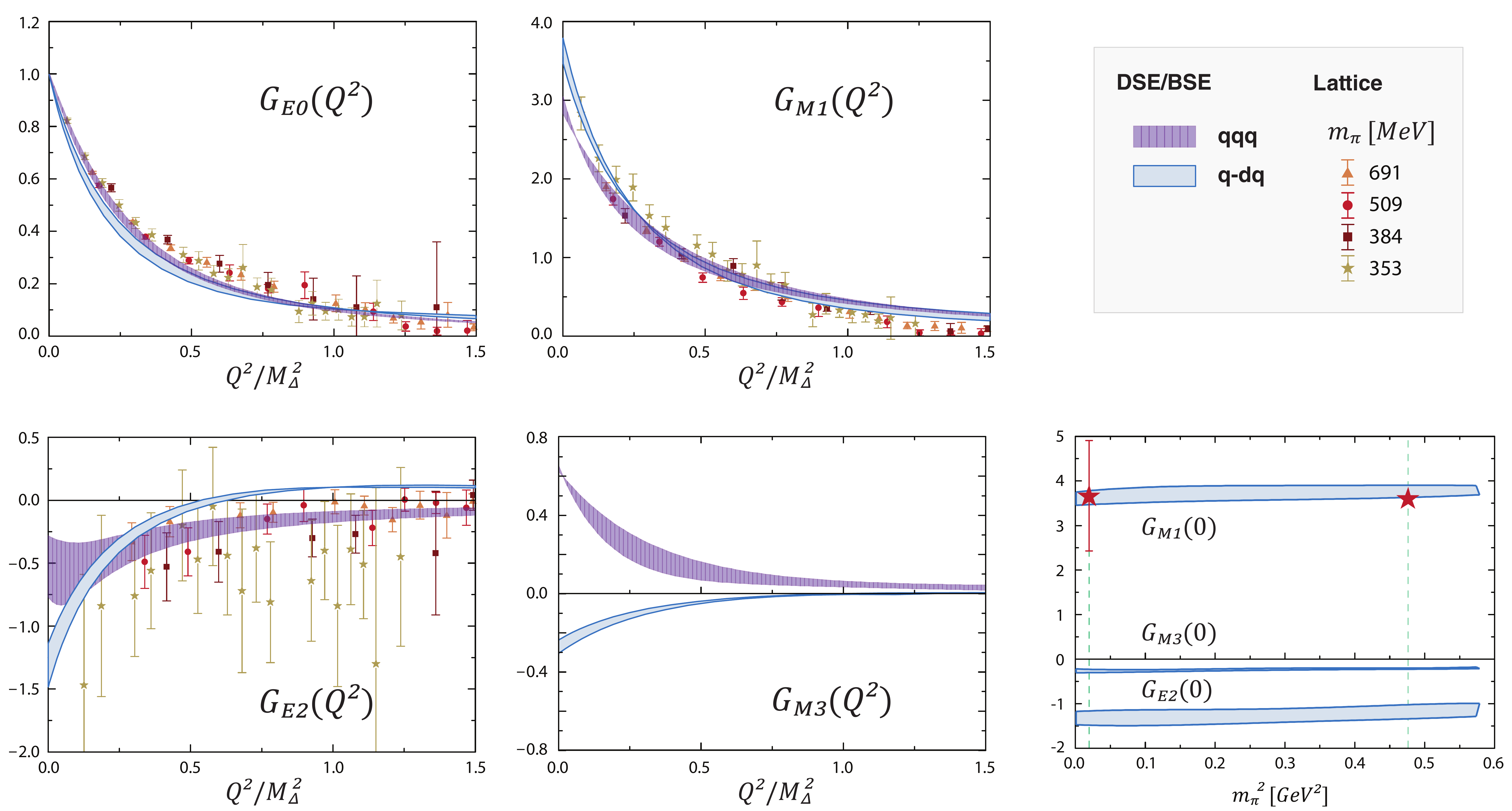}
                    \caption{Electric monopole, magnetic dipole, electric quadrupole
and magnetic octupole form factors for the $\Delta^+$ baryon. We compare lattice data~\cite{Alexandrou:2009hs}
with the Dyson-Schwinger three-quark~\cite{Sanchis-Alepuz:2013iia} and quark-diquark calculations~\cite{Nicmorus:2010sd}.
The bands for the DSE results reflect the systematic uncertainty.
The plot on the bottom right shows the current-mass evolution of the static quantities from the quark-diquark calculation~\cite{Nicmorus:2010sd},
where stars denote the experimental magnetic moments of the $\Delta^+$ and $\Omega^-$ baryon. (The value $m_\pi^2 \approx 0.47$ where the magnetic 
moment of the all-strange $\Omega^-$ baryon is plotted corresponds to a pion mass evaluated with constituent strange quarks.)} \label{fig:delta-ffs}
        \end{figure*}

Also very encouraging are the results for the strange baryon sector reported for the three-body framework in~\cite{Sanchis-Alepuz:2014sca}.
Here the masses of ground-state octet and decuplet baryons have been reproduced on the few-percent level, while the calculation
of excited states still needs to be performed.

\section{Selected form factors: electromagnetic $\Delta$ form factor}\label{sec-4}

Form factors in the functional approach are calculated by coupling external currents to equations for Green functions.
In the three-body approach, electromagnetic form factors~\cite{Eichmann:2011vu},
axial form factors~\cite{Eichmann:2011pv} and hyperon form factors~\cite{Sanchis-Alepuz:2015fcg} have been calculated and
reviewed in~\cite{Eichmann:2016yit}. Here we discuss aspects of the electromagnetic form factors for the $\Delta$, which have been
determined in the three-body approach~\cite{Sanchis-Alepuz:2013iia} as well as in the quark-diquark approximation~\cite{Nicmorus:2010sd}.
In Fig.~\ref{fig:delta-ffs} we display corresponding results for the electric monopole and quadrupole form factor as well as the magnetic dipole and octupole form factor.
We compare them with
results from lattice QCD~\cite{Alexandrou:2009hs} using different values for the up/down quark masses corresponding to different
pion masses.

It is interesting to compare the three-body with the quark-diquark results. For the
electric monopole form factor both approaches agree very well. The limit of vanishing photon momentum, $Q^2 \rightarrow 0$,
is constrained by charge conservation,
which is guaranteed automatically in our framework due to gauge invariance. For all $Q^2$ values the agreement with lattice
results is furthermore on a quantitative level. The electric quadrupole form factor, however, is more sensitive to the internal
structure of the $\Delta$ baryon, as can be seen in the bottom left diagram of Fig.~\ref{fig:delta-ffs}. Here the results from the
three-body and the quark-diquark approach display a different momentum behaviour and even a different sign at large momenta.
The same signature is found for the magnetic form factors. While the dipole results agree with each other (apart from a small
region at very low $Q^2$), the octupole component has a different sign for all $Q^2$.
Thus, high precision lattice results are clearly able to distinguish between them. Given the presently available set of data
for the electric quadrupole component,
there may be a slight tendency towards the three-body results. However, the accuracy is not good enough to make a definite statement.
Thus, it would be very interesting to repeat the lattice calculation with today's technical and numerical resources.

While experimental results for the $\Delta$ form factors are not available, static quantities like its magnetic moment can be extracted.
These are shown in the bottom right plot of Fig.~\ref{fig:delta-ffs}. Compared are the bands for the quark-diquark results with the
experimental data for the $\Delta^+$ and the $\Omega^-$ baryon. The agreement with the central value of the experimental error bar
is very good. Note that the magnetic moments are real, since in the rainbow-ladder approximation used in the present approach decay channels
are not included, see the review Ref.~\cite{Eichmann:2016yit} for a more detailed discussion of this aspect. As discussed before, the magnetic 
moment of the $\Delta^+$ extracted from the magnetic dipole form factor shows a small discrepancy between the three-body and the 
quark-diquark results. Thus also more precise experimental results have the potential to discriminate between the two.

In general, we confirm the $\Delta$ baryon's deformation as encoded in its electric quadrupole and magnetic
octupole form factors. Non-relativistically, a negative sign for the electric quadrupole moment $G_{E2}(0)$
indicates an oblate charge distribution for the $\Delta$ in the Breit frame~\cite{Alexandrou:2009hs}.
From the measurement of the $\gamma N\to\Delta$ transition one can infer the value $G_{E2}(0)=-1.87(8)$ in the large-$N_C$ limit~\cite{Buchmann:2002mm,Alexandrou:2009hs}; comparable values are predicted by a range of
constituent-quark models~\cite{Ramalho:2009vc}. These numbers are in the ballpark of the lattice and DSE results
displayed in Fig.~\ref{fig:delta-ffs}.

\section{Conclusions}\label{sec-5}
Baryon spectroscopy, baryon structure and baryon dynamics remain very active
and lively fields. Dedicated experiments at many facilities such as Jefferson Lab,
BESIII, ELSA, J-PARC, LHCb, MAMI and the future PANDA/FAIR experiment are contributing
and will continue to contribute to our understanding of the baryonic world.
On the theory side, the past years have seen continuous advances in connecting the
underlying quark and gluon interactions of QCD with baryon phenomenology.

In functional approaches, the challenges are to capitalize on continuous improvements
to the truncation/approximation schemes. While the aim is not to enter the realm
of high precision physics on the sub-percent level, much can be achieved in the way
of understanding the underlying mechanisms that generate the observed phenomena.
A prime example of this is the question of three-quark vs.\ a quark-diquark structure of
baryons. At face value, the results reported in this contribution
suggest that irreducible three-body forces inside baryons may be subdominant
compared to two-body interactions. If this is the case, the quark-diquark approximation
should be fine. Indeed, accounting for well-known deficiencies in the rainbow-ladder
truncation of the quark-quark interaction, we found a spectrum of ground and excited
states of light baryons in one-to-one agreement with experiment. This highly non-trivial
result is very encouraging and suggests further extensions of our work to the strange
baryon sector and quantum numbers beyond $J=\tfrac{1}{2}$ and $J=\tfrac{3}{2}$.

On the other hand, we have seen that form factors may serve to distinguish between a
three-body and a quark-diquark picture of baryons via static quantities at low momenta
and the dynamical behaviour at larger momenta. While the former can be tested by more
precise experiments, the latter can be accessed by lattice QCD. 
\vspace*{2mm}

\begin{acknowledgement}
{\bf Acknowledgement}\\
The work reported here would not have been possible without the
support from the Deutsche Forschungsgemeinschaft within the SFB/TR16.
I thank the members of SFB/TR16 for many fruitful discussions and the
spokesperson Ulrike Thoma for the excellent leadership.
I am furthermore very grateful to Gernot Eichmann, Helios Sanchis-Alepuz 
and Richard Williams for the successful and pleasant collaboration on the
topics summarized in this proceedings contribution within the framework 
of the SFB/TR16. In addition I would like to thank Reinhard Alkofer (together 
with the above mentioned gentlemen) for many discussions and the collaboration 
in the production of an extensive review article \cite{Eichmann:2016yit} 
on the topics of the SFB/TR16.
\end{acknowledgement}

\bibliography{baryons}


\end{document}